\documentclass[journal,12pt, onecolumn]{IEEEtran}

\IEEEoverridecommandlockouts

\usepackage{subfigure}
\usepackage{pdfpages}
\usepackage{diagbox}
\usepackage{setspace}

\usepackage{array}
\usepackage{caption}

\usepackage{amsmath,bm}
\usepackage{cite}
\usepackage{amsmath,amssymb,amsfonts}
\usepackage{algorithmic}
\usepackage{amsthm}
\usepackage{graphicx}
\usepackage{textcomp}
\usepackage{xcolor}
\def\BibTeX{{\rm B\kern-.05em{\sc i\kern-.025em b}\kern-.08em
    T\kern-.1667em\lower.7ex\hbox{E}\kern-.125emX}}
\begin{document}
\begin{spacing}{1.55}

\title{Computation Offloading in Multi-Access Edge Computing Networks: A Multi-Task Learning Approach  
}

\author{
\IEEEauthorblockN{Bo Yang\IEEEauthorrefmark{1}, Xuelin Cao\IEEEauthorrefmark{2}, Joshua Bassey\IEEEauthorrefmark{1},  Xiangfang Li\IEEEauthorrefmark{1}, Timothy~Kroecker\IEEEauthorrefmark{3}, Lijun Qian\IEEEauthorrefmark{1}} 
  \small
\IEEEauthorblockA{\IEEEauthorrefmark{1}Department of Electrical and Computer Engineering and CREDIT Center, Prairie View A$\&$M University, \\
Texas A$\&$M University System, Prairie View, TX 77446, USA} 

\IEEEauthorblockA{\IEEEauthorrefmark{2}Department of Electrical and Computer Engineering, University of Houston, Houston, TX 77004, USA}

\IEEEauthorblockA{\IEEEauthorrefmark{3}US Air Force Research Laboratory (AFRL), Rome, NY 13441, USA}
}


\maketitle

\begin{abstract}
Multi-access edge computing (MEC) has already shown the potential in enabling mobile devices to bear the computation-intensive applications by offloading some tasks to a nearby access point (AP) integrated with a MEC server (MES). However, due to the varying network conditions and limited computation resources of the MES, the offloading decisions taken by a mobile device and the computational resources allocated by the MES  may not be efficiently achieved with the lowest cost.  In this paper, we propose a dynamic offloading framework for the MEC network, in which the uplink non-orthogonal multiple access (NOMA) is used to enable multiple devices to upload their tasks via the same frequency band. We formulate the offloading decision problem as a multiclass classification problem and formulate the MES computational resource allocation problem as a regression problem. Then a multi-task learning based feedforward neural network (MTFNN) model is designed to jointly optimize the offloading decision and computational resource allocation.  Numerical results illustrate that the proposed MTFNN outperforms the conventional optimization method in terms of inference accuracy and computation complexity.
\end{abstract}

\section{Introduction}
\begin{figure}[t]
  \centering
  \captionsetup{font={small }}
  \includegraphics[width=2.4in, height=1.9in]{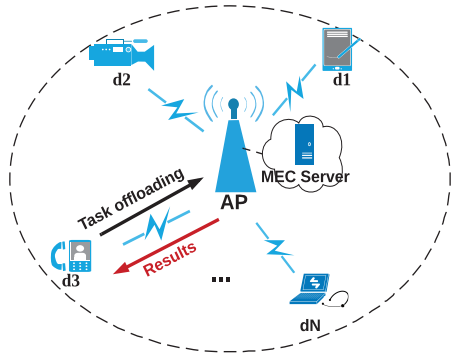}
  \caption{Task offloading scenario in MEC system}
  \label{fig01}
\end{figure}

To cope with the exponentially increasing data traffic with stringent requirements on computation resources, multi-access edge computing (MEC) network plays a key role in bringing cloud functionalities to the edge that in close proximity to mobile devices which support multiple access\cite{survey}. With computation offloading technique, the resource-constrained mobile devices can save energy and enrich users' experience by fully or partially offloading computation-intensive tasks to the nearby MEC server (MES). The MES could be colocated with access points (APs), wireless relays or small base stations (BSs), as shown in Fig.~\ref{fig01}. Due to a large amount of computation input data may be uploaded from the mobile devices to the MES, abundant wireless spectrum is required, which has become more and more scarce and precious. To alleviate this problem, non-orthogonal multiple access (NOMA) has been introduced into the (MEC) networks of the 5G era enabling multiple devices to transmit their data simultaneously on the same frequency band\cite{NOMA}. 

In order to minimize the tasks completion time and energy consumption of the mobile devices, one of the challenges in offloading computation-intensive tasks to MES is to \emph{determine whether to offload and the portion of computational resources allocated to the device}. This could be formulated as a mixed integer nonlinear programming (MINLP) problem that minimizes the total system cost (e.g., the weighted sum of delay and energy consumption) under the constraints of the task's tolerable delay and MES's available resources, which is NP-hard in general~\cite{MINLP-1}.  Furthermore, \emph{the input parameters to the optimization problem may vary frequently, which leads to the requirement of the offload decision making in near-real-time}. 
However, the conventional optimization algorithms usually take a lot of time to solve this NP-hard optimization problem and only sub-optimal solutions are obtained in some cases. As a result, the optimal or sub-optimal offloading decision may not be reached within the specified delay constraints (e.g., in the video conference, real-time image processing, etc.). 

In order to address this challenge, a novel computation offloading framework is proposed in this work that can adapt to the varying network conditions and the requirements of devices' applications. 
Specifically, a multi-task learning based feedforward neural network (MTFNN) model is designed to solve the mixed integer nonlinear programming (MINLP) problem in near-real-time, where the offloading decision making is formulated as a multiclass classification problem and the computational resource allocation is formulated as a regression problem. The proposed MTFNN model is firstly trained offline with the dataset collected by traversing all the possible combinations of features including the parameters representing wireless channel conditions. Then the trained MTFFN model can be used to predict the optimal offloading decision and computational resource allocation in near-real-time with high accuracy.  Simulation results show that the proposed MTFFN model based offloading scheme achieves better performance compared with existing benchmark offloading approaches.

The remainder of this paper is organized as follows. Section II presents the system models. In Section III, we present the formulation and analysis of the cost minimization problem. In Section IV, we describe our proposed offloading scheme in detail. Numerical results are given in Section V, followed by a review of related works in Section VI. Finally, we conclude this paper in Section VII.

\section{System Models}
\subsection{Network Model}
We consider a scenario of multi-devices single-MES in the MEC network, as shown in Fig. \ref{fig01}. There exist $N$ devices, i.e., ${\cal N}=\{d_1, d_2, ..., d_N\}$, which are associated with the MES located in the AP (denoted as $\cal A$). In this architecture, the widely deployed WLAN can be considered as a potential technology for wireless communications, which works on the unlicensed frequency band. We assume that each device $d_i, \forall i  \in[1, N]$, has only one computation-intensive task (denoted as ${\cal J}_i$) to be processed during a computation offloading period, which is atomic and cannot be further divided. Each device can choose to offload ${\cal J}_i$ to the MES through wireless links or execute it locally In general, the total computation ability and storage capacity of the MES is limited and thus maybe not always sufficient for all associated devices to offload their tasks simultaneously. 


\newtheorem{theorem}{\textit{Theorem}}
\newtheorem{lemma}{\textit{Lemma}}
\subsection{Task Model}
For ${\cal J}_i, \forall i \in[1,N]$, $d_i$ can only execute it locally or by offloading computing in the MES. Denote $D_i\in \{0,1\}$ as the computation offloading decision, which is a $N$-dimensional binary vector (denoted as $\mathbf{D}$), i.e., $\mathbf{D}=[D_1, D_2,...,D_N]$. Specifically, we have $D_i=0$ when $d_i$ executes ${\cal J}_i$ locally. Otherwise, $D_i=1$ can hold for the computation offloading. Moreover, we let $\mathbf{F}=[f_1, f_2,...,f_N]$ be the allocated computational resource (i.e., central processing unit (CPU) cycles per second) vector by the MES. Hence, the offloading strategy can be defined as $\bm{S}=\{\mathbf{D}, \mathbf{F}\}$. In order to make ${\cal J}_i$ more visible and intuitive, we characterize ${\cal J}_i$ by a three-tuple of parameters, i.e., ${\cal J}_i(s_i, c_i, \vartheta_i)$. In particular,  $s_i$ denotes the size of computation input data needed for processing ${\cal J}_i$. $c_i$ denotes the total number of CPU cycles required to process ${\cal J}_i$, and $\vartheta_i$ denotes the maximum tolerable delay of ${\cal J}_i$.
 
\subsection{Communication Model}
In the uplink NOMA system, the received signal from $d_i$ in $\cal A$ is given as
\begin{equation}\label{e01}
y_i=\underbrace{\sqrt{P_t^i}h_ix_s}_{\rm Desired \ signal}+\underbrace{\sum_{j\neq i, j\in{\cal N}}{\sqrt{P_t^j}h_jx_j}}_{\rm Interferences}+\underbrace{z_i}_{\rm Noise},
\end{equation}
where $h_i$ denotes the channel power gain\footnote{It is assumed that the channel remains static within each time frame, in which the optimal offloading strategy $\bm{S}=\{\mathbf{D}^*, \mathbf{F}^*\}$ can be obtained.} for $d_i$ connecting with $\cal A$. $P_t^i$ is the transmit power of $d_i$, and the noise power $z_i$ can be generally considered as the white Gaussian noise in additive white Gaussian noise (AWGN) channel with zero mean and variance $\delta^2$. 

To separate and decode the overlapped signals from $d_i, \forall i \in [1,N]$, the successive interference cancellation (SIC) modular can be implemented in $\cal A$. By sorting the overlapped signals descendly according to the channel gains, i.e.,
\begin{equation}\label{e02}
h_1 \geqslant h_2 \geqslant h_3 \geqslant ... \geqslant h_N, \ \forall i\in [1, N],
\end{equation}
the received signal-to-interference-plus-noise ratio (SINR) of $d_i$ served by $\cal A$ can be calculated as
\begin{equation}\label{e03}
{\rm{SINR}}_i=\frac{P_t^i |h_i|^2}{\delta^2+\sum_{j=i+1}^{N}P_t^j |h_j|^2}. 
\end{equation}

\subsection{Computation Model}
For the offloading strategy $\bm{S}=\{\mathbf{D}, \mathbf{F}\}$, the offloading decision of $d_i,  \forall i \in [1,N]$ can be ``locally" or ``offloading", i.e., $D_i\in \{0,1\}$, so the two computation models are presented as follows.
\subsubsection{Locally}
Let $\tau _{l}^{i}$ be the local execution delay of ${\cal J}_i$, denote $f _{l}^{i}$ as the the CPU cycle frequency (i.e., CPU cycles per second) of $d_i$\footnote{Without loss generality, we assume that the computational capabilities of each device may be different.}. The local execution delay of ${\cal J}_i$ is
\begin{equation}\label{e04}
\tau _{l}^{i}=\frac{c_i}{f{_{l}^{i}}}.
\end{equation}
 
Denote $\kappa$ as the energy efficiency parameter that is mainly depends on the chip architecture\cite{k}. In order to process ${\cal J}_i$ with the CPU clock speed $f _{l}^{i}$, the energy consumption is
\begin{equation}\label{e05}
\varepsilon_{l}^{i}=\kappa \left ( f_{l}^{i} \right )^2 c_i.
\end{equation}

Based on (\ref{e04}), (\ref{e05}), the total cost for computing ${\cal J}_i$ locally is
\begin{equation}\label{e06}
{\cal O} _{l}^{i}=\alpha  \tau _{l}^{i} + \beta  \varepsilon_{l}^{i},
\end{equation}
where $\alpha$ and $\beta$ are the weights of execution delay and energy consumption. In general, $0 \le \alpha, \beta \le 1$ and $\alpha+\beta=1$ hold.  

\subsubsection{Offloading}
To process ${\cal J}_i$ with the offloading approach, $d_i$ firstly needs to upload the data to $\cal A$ which is co-located with the MES through the wireless access network. Then, the MES allocates the computational resources accordingly and execute ${\cal J}_i$ instead. Finally, $\cal A$ returns the executing results to $d_i$. In the following, we describe the three stages in detail.

\begin{itemize}
\item \textit{Uploading}. The delay to upload data to $\cal A$ is
\begin{equation}\label{e07}
T_{u}^{i}= \frac{s_i}{r_{u}^{i}},
\end{equation}
where $r_{u}^{i}$ stands for the achieved uplink data rate of the wireless link from $d_i$ to $\cal A$. Denote $W$ as the frequency bandwidth, we have $r_{u}^{i}=W log_2{\left ( 1+ {\rm{SINR}}_i \right )}$.

The energy consumption during the data uploading is 
\begin{equation}\label{e08}
e_{u}^{i}=P_{t}^{i} T_{u}^{i}=\frac{P_{t}^{i}s_i}{W log_2{\left ( 1+ {\rm{SINR}}_i \right )}}.
\end{equation}

\item \textit{Processing}. The time to process ${\cal J}_i$ by the MES is 
\begin{equation}\label{e10}
T_{p}^{i}=\frac{c_i}{f_i},
\end{equation}
where $f_i$ denotes the allocated computational resource to $d_i$ by the MES. Let $F$ be the entire resources of the MES, we have $\sum_{i=1}^{N}D_i f_i\leq F$.

We suppose that $d_i$ stays idle while waiting for the results from MES, the power consumption is defined as $P_{I}^{i}$. The energy consumption is
\begin{equation}\label{e11}
e_{I}^{i}=P_{I}^{i} T_{p}^{i}=\frac{P_{I}^{i} c_i}{f_i}.
\end{equation}

\item \textit{Downloading}. The time to download the executive results from $\cal A$ is
\begin{equation}\label{e12}
T_{d}^{i}=\frac{w_i}{r_{d}^{i}},
\end{equation}
where $w_i$ is the size of the results, $r_{d}^{i}$ denotes the data rate of the wireless down link between $\cal A$ and $d_i$.

For $d_i, \ \forall i \in [1,N]$, denote the power required to download the execusive results as $P_{d}^{i}$. 

Accordingly, the energy consumption of $d_i$ during downloading the results is
\begin{equation}\label{e13}
e_{d}^{i}=P_{d}^{i} T_{d}^{i}=\frac{P_{d}^{i} w_i}{r_{d}^{i}}.
\end{equation}
\end{itemize}

Generally, due to $w_i\ll s_i$ (e.g., face recognition) and $r_{d}^{i}$ is relatively high, the total execution delay and energy consumption of $d_i$ can be approximately given as
\begin{equation}\label{e14}
\tau_{o}^{i}\approx \frac{s_i}{W  log_2{\left ( 1+ {\rm SINR}_i \right )}}+\frac{c_i}{f_i},
\end{equation}
\begin{equation}\label{e15}
\varepsilon_{o}^{i}\approx \frac{P_{t}^{i}s_i}{W log_2{\left ( 1+ {\rm{SINR}}_i \right )}} + \frac{P_{I}^{i} c_i}{f_i},
\end{equation}
where $T_{d}^{i}$ and $e_{d}^{i}$ can be neglected\cite{TDMA}.

Therefore, the total cost for computing ${\cal J}_i$ is
\begin{equation}\label{e16}
{\cal O}_{o}^{i}=\alpha  \tau_{o}^{i} + \beta \varepsilon_{o}^{i}.
\end{equation}

To this end, the sum cost of all devices can be expressed as 
\begin{equation}\label{e17}
{\cal O}_{total}=\sum_{i=1}^{N} \left ({1-D_i}\right ) {\cal O}_{l}^{i}+ D_i  {\cal O}_{o}^{i}.
\end{equation}

\section{Cost Minimization Problem}
\subsection{Problem Formulation}
In this subsection, we formulate the offloading and resource allocation by the MES as a cost minimization problem ($\mathbf{P1}$). 
\begin{subequations} \label{e18}
\begin{align}
&\mathbf{P1}:\;\; \underset{\mathbf{D, F}}{\rm minimize}\;\; {\cal O}_{total} \notag \\
& \;\;\;\;\;\;{\rm{s}}{\rm{.t}}{\rm{.}}\;\;\;\mathbf{C1}: \; D_i\in \left \{0,1  \right \}, \ \forall i \in N, \\
& \;\;\;\;\;\;\;\;\;\;\;\;\;\;\mathbf{C2}: \; \left ({1-D_i}\right ) { \tau}_{l}^{i}+ D_i { \tau}_{o}^{i}  \le  \vartheta_i, \\
& \;\;\;\;\;\;\;\;\;\;\;\;\;\;\mathbf{C3}: \; 0 \le f_i \le F, \ \forall i \in N, \\
& \;\;\;\;\;\;\;\;\;\;\;\;\;\;\mathbf{C4}: \; \sum_{i=1}^{N}D_i f_i \le F, \ \forall i \in N.
 \end{align}
\end{subequations}

In the optimization problem, $\mathbf{D} = [D_1, D_2,...,D_N]$ is the offloading decision, and $\mathbf{F} = [f_1, f_2,...,f_N]$ denotes the computational resource allocation. $\mathbf{C1}$ shows that $d_i$ can only choose to execute ${\cal J}_i$ locally or offloading to the MES. $\mathbf{C2}$ makes sure that the time cost to process ${\cal J}_i$ should not exceed the maximum tolerable delay $\vartheta_i$. $\mathbf{C3}$ and $\mathbf{C4}$ guarantee that the computational resource allocated to $d_i$ and the sum of the computational resources allocated to all the offloading devices should not exceed the total resources of the MES. 

\subsection{Problem Analysis}
Intuitively, the optimization problem $ \mathbf{P1}$ can be solved by going through all the combinations of the offloading decision vector $\mathbf{D}$ and the computational resource allocation $\mathbf{F}$. Denote the optimal offloading decision and computational resource allocation result as $\mathbf{D}^*$ and $\mathbf{F}^*$, i.e.,
\begin{equation}\label{e19}
\left \{ {\mathbf{D}^*, \mathbf{F}^*} \right \}=\underset{\mathbf{D, F}}{{\rm argmin}}\ {\cal O}_{total}.
\end{equation}
 
However, due to the fact that $\mathbf{D}$ is the binary vector, and the objective function of $\mathbf{P1}$ is not convex, so the resolving of $\mathbf{P1}$ is difficult to tackle \cite{MINLP-1}. Generally, the spatial branch and bound (sBB) method is used to solve this problem \cite{MINLP-2}, where a hierarchy of nodes represented by a binary tree is created (a.k.a. the sBB tree) and then a pure continuous NLP sub-problem can be formed by dropping the integrality requirements of the discrete variables~\cite{MINLP-3}. As a result, the initial optimization problem $\mathbf{P1}$ becomes the root of the sBB tree.
Although the sBB can resolve the MINLP problem faster than the exhaustive searching, large overhead will be still introduced into the MEC networks due to the varying of channel condition and input parameters. Moreover, the obtained results using the sBB method are sometimes sub-optimal, which degrades the performance of the MEC system. In this paper, instead of the conventional optimization methods, we build a machine learning model to predict $\mathbf{D}^*$ and $\mathbf{F}^*$ more efficiently while ensuring the prediction accuracy.

\section{Computation Offloading with Multi-task Learning}
\subsection{Problem Mapping}
\begin{figure}[t]
  \centering
  \captionsetup{font={small }}
  \includegraphics[width=4.5in, height=2.55in]{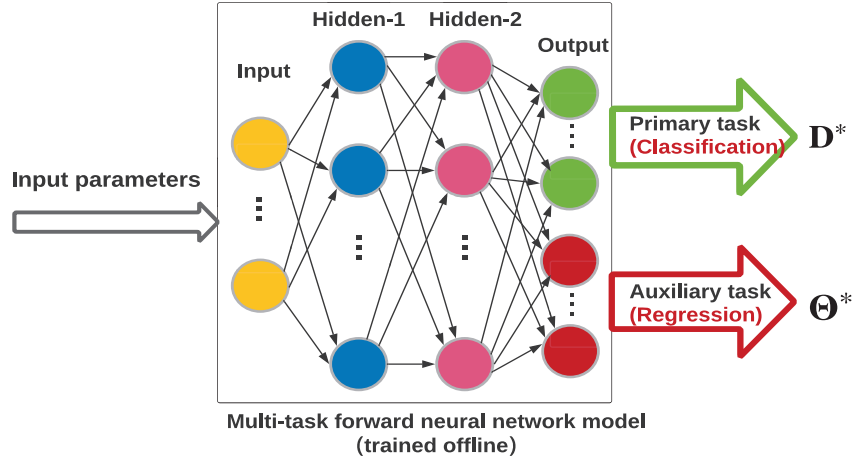}
  \caption{The proposed MTFNN model with an input layer, two hidden layers, and an output layer. Taking $N=3$ as an example, the input contains $18$ neurons, the two hidden layers contain $15$ and $10$ neurons, respectively. In the output layer, the classification output contains $8$ neurons and the regression output contains $3$ neurons}
  \label{fig02}
\end{figure}

The two output vectors (i.e., $\mathbf{D}^*$ and $\mathbf{F}^*$) of $\mathbf{P1}$ are related to each other. If we consider the prediction of $\mathbf{D}^*$ and $\mathbf{F}^*$ as two machine learning tasks, it is known that learning the two related tasks jointly can get better generalization effect than the learning them individually \cite{MTL}. Therefore, $\mathbf{P1}$ can be formulated as a multi-task learning (MTL) problem, as shown in Fig. \ref{fig02}, where the MTFNN model is proposed to predict  $\mathbf{D}^*$ and $\mathbf{F}^*$ with multi-task learning. Suppose that there exist $l$ learning tasks $\left \{ {\cal T}_i \right\}{_{i=1}^{l}}$ that are related to each other, where $l=2$ in our proposed MTFNN model. Each learning task ${\cal T}_i$ is usually accompanied by a training dataset ${\cal S}_i$ which consists  of $m_i$ training samples, i.e., ${\cal S}_i = \left \{ \mathbf{X}_{j}^{(i)},\mathbf{Y}_{j}^{(i)} \right \}_{j=1}^{m_i}$, where $\mathbf{X}_{j}^{(i)}$ is the $j$-th training instance in ${\cal T}_i$, $\mathbf{Y}_{j}^{(i)}$ represents its label.  For the output, denote $y_{j}^{(i)}$ as the $j$-th corresponding output from $\mathbf{Y}_{j}^{(i)}$. When $y_{j}^{(i)}$ is in a discrete space, e.g., $y_{j}^{(i)}\in \left \{ 0,1 \right \}$ for $\mathbf{D}^*$, and thus the corresponding task can be considered as a multiclass classification problem, where the task is to predict a discrete offloading decision class for a given set of input parameters. If $y_{j}^{(i)}$ is continuous, e.g., $y_{j}^{(i)}\in \mathbb{R}$ for $\mathbf{F}^*$, the corresponding task turns to be a regression problem, where the task is to predict a numeric value. It should be noted that in our regression model, we define all the labels as $\mathbf{\Theta}_{j}^{(i)} = \mathbf{Y}_{j}^{(i)}/F$ for simplicity. Therefore, instead of $\mathbf{F}$, the prediction of the regression model becomes a computational resource allocation ratio, i.e., $\mathbf{\Theta}_{j}^{(i)} \in [0.0,1.0]$. Owing that the offloading decision plays a more important role in the MEC networks, so the prediction of $\mathbf{D}^*$ and $\mathbf{\Theta}^*$ can be considered as a primary task (a.k.a. classification problem) and a auxiliary task (a.k.a. regression problem) in our MTFNN model, respectively. 

\subsection{Data Collection}
We independently generate $4\times 10^4$, $5\times 10^4$, $8\times 10^4$ and $10^5$ data samples for $N \in [2, 5]$ in the dataset by traversing all the possible combinations of $\mathbf D$ and $\mathbf \Theta$ with the exhaustive searching algorithm\footnote{Due to the $\mathbf{\Theta}$ is a decimal vector which ranges from $[0.0, \ 1.0]$. In this paper, the interval between traversal values is set as $0.1$, which is also denoted as the granularity of resource allocation ($\omega$) in the follow-up contents.}, so ${\mathbf D}^*$ and ${\mathbf \Theta}^*$ can be obtained for a given set of parameters. During each execution, the network parameters are randomly chosen from their ranges given in Table I, and the statical parameters are given as follows. The channel bandwidth ($W$) is $1$ MHz, and the white noise power is ($\delta^2$) is $7.9\times10^{-13}$. The energy efficiency parameter ($\kappa$) is set as $1\times10^{-28}$. The CPU computation capacity of the MES ($F$) is $2.5$ GHz. The transmission power ($P_t$) and idle power ($P_I$) of each device are set to be $0.3$ W and $0.1$ W, respectively\cite{power}. The uplink data rate ($r_{u}^{i}$) can be calculated according to $r_{u}^{i}=W log_2{\left ( 1+ {\rm{SINR}}_i \right )}$. In order to enable the collected data to be applied to our MTFNN model, we preprocess the dataset as a specific groundtruth matrix $\mathbf H$. Specifically, for each device $d_i, \forall i \in [1, N]$, the input parameters of the MTFNN model include $s_i$, $c_i$, $f_l^i$, $h_i$, $\alpha_i$ and $\beta_i$. The output from the MTFNN model includes ${\mathbf D}^*$ and ${\mathbf \Theta}^*$. The  collected dataset is split into $80\%$ for training phase and the rest $20\%$ for testing phase.

\begin{table}[t]  
\renewcommand{\arraystretch}{1.2}
  \captionsetup{font={small}} 
\caption{\\ \scshape Critical Parameters and Definitions } 
\label{notations}  
\centering  
\begin{tabular}{l | l}  
\hline
Parameters & Value range \\
\hline 
 The number of devices ($N$) & $2,3,4, 5$ \\
 Data payload size ($s$) & $[1 - 500]$ kbits\\
 CPU cycle required to process the data ($c$) & $[3 - 1500]$ Megacycles \\
 CPU frequency of the device ($f_l$) &  $[1 {\rm Hz} -  1 {\rm GHz}]$ \\
 Weights of delay and energy cost ($\alpha$, $\beta$) & $[0.0 - 1.0]$ \\
  \hline
\end{tabular}  
\end{table} 

\begin{figure*}[t] 
\centering
\captionsetup{font={footnotesize }}
\subfigure[]{
\label{Fig-06-1}
\includegraphics[width=2.2in,height=1.95in]{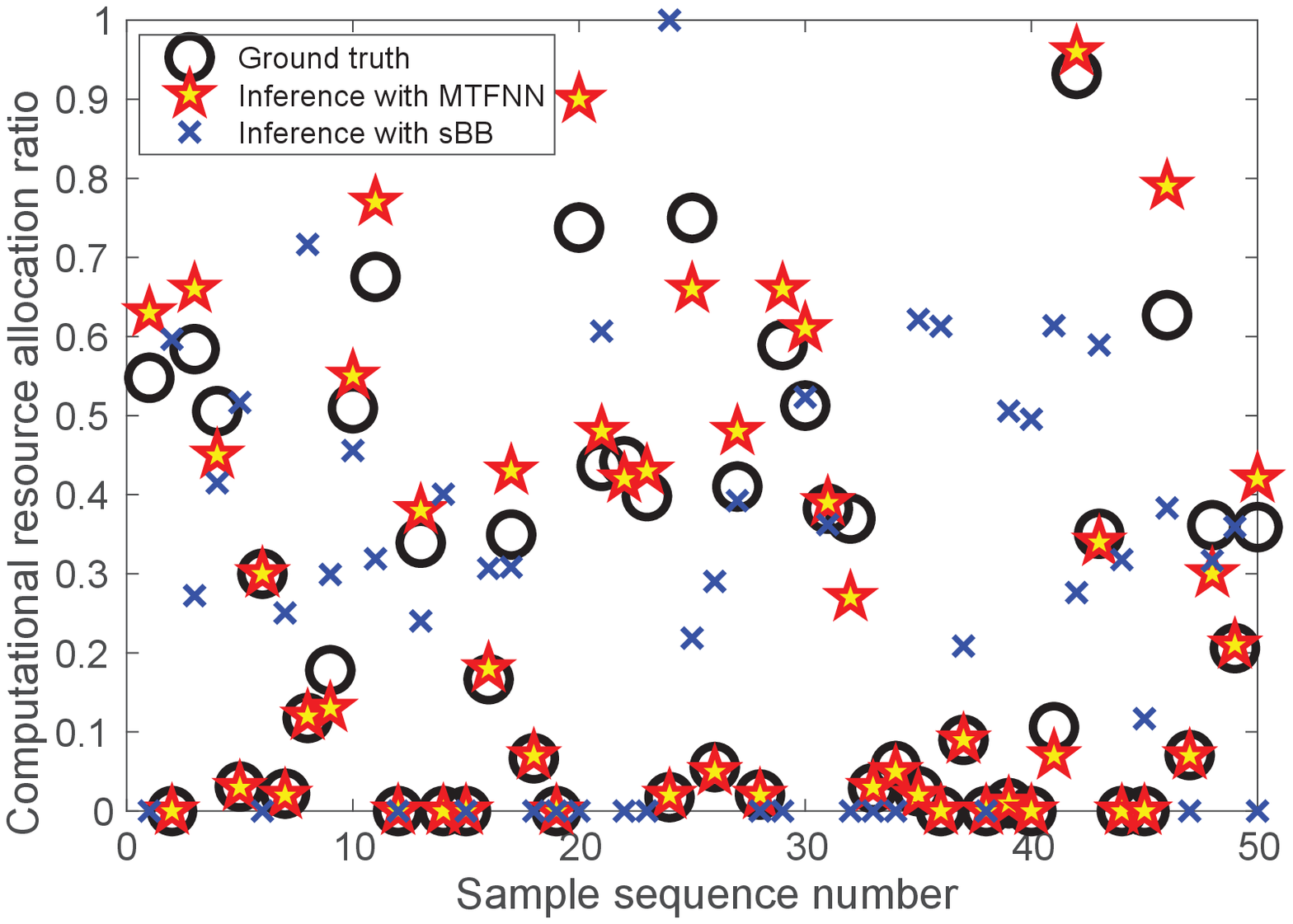}} 
\hspace{-0.05in}
\subfigure[]{
\label{Fig-06-2}
\includegraphics[width=2.2in,height=1.95in]{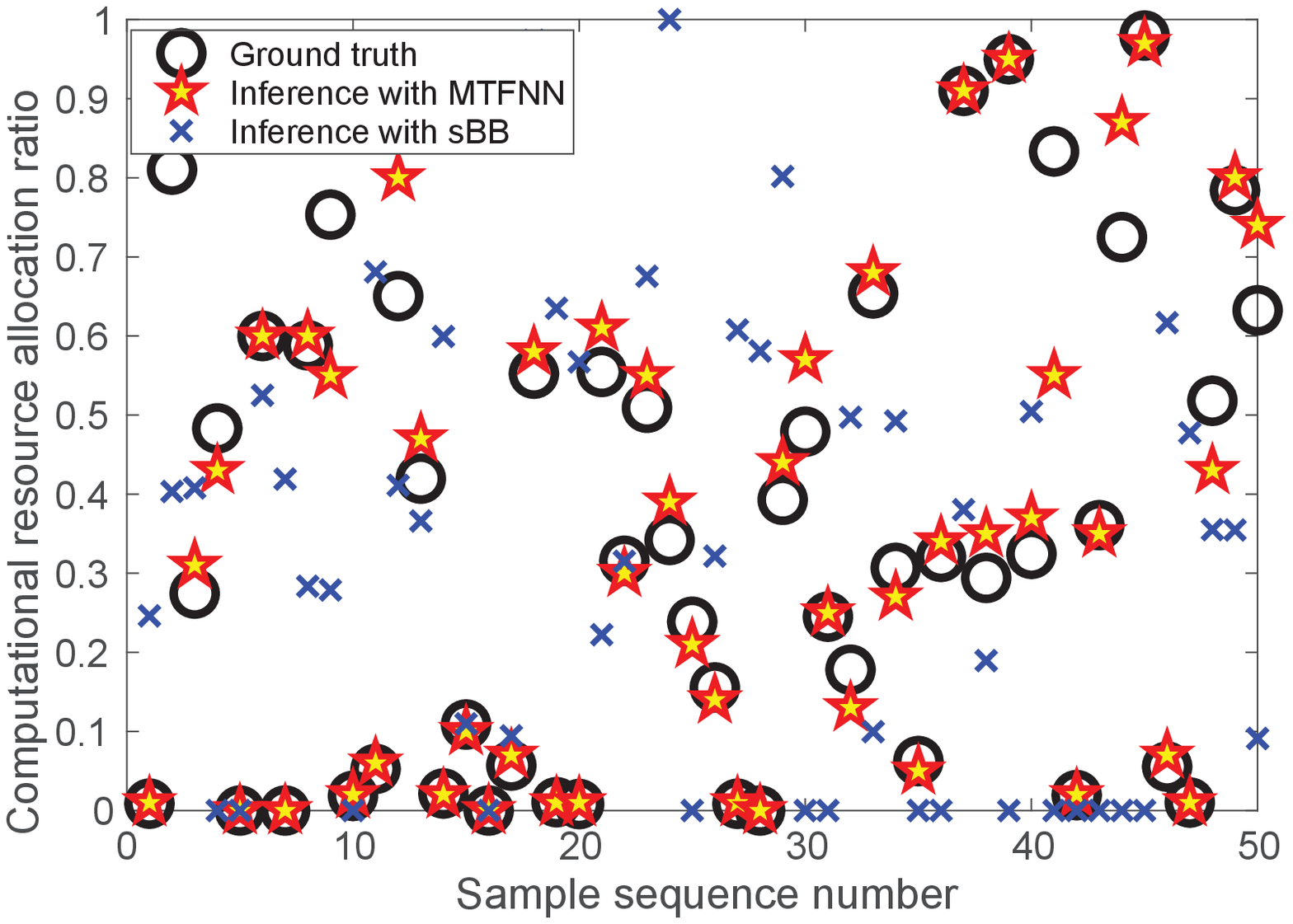}}
\hspace{-0.05in}
\subfigure[]{
\label{Fig-06-2}
\includegraphics[width=2.2in,height=1.95in]{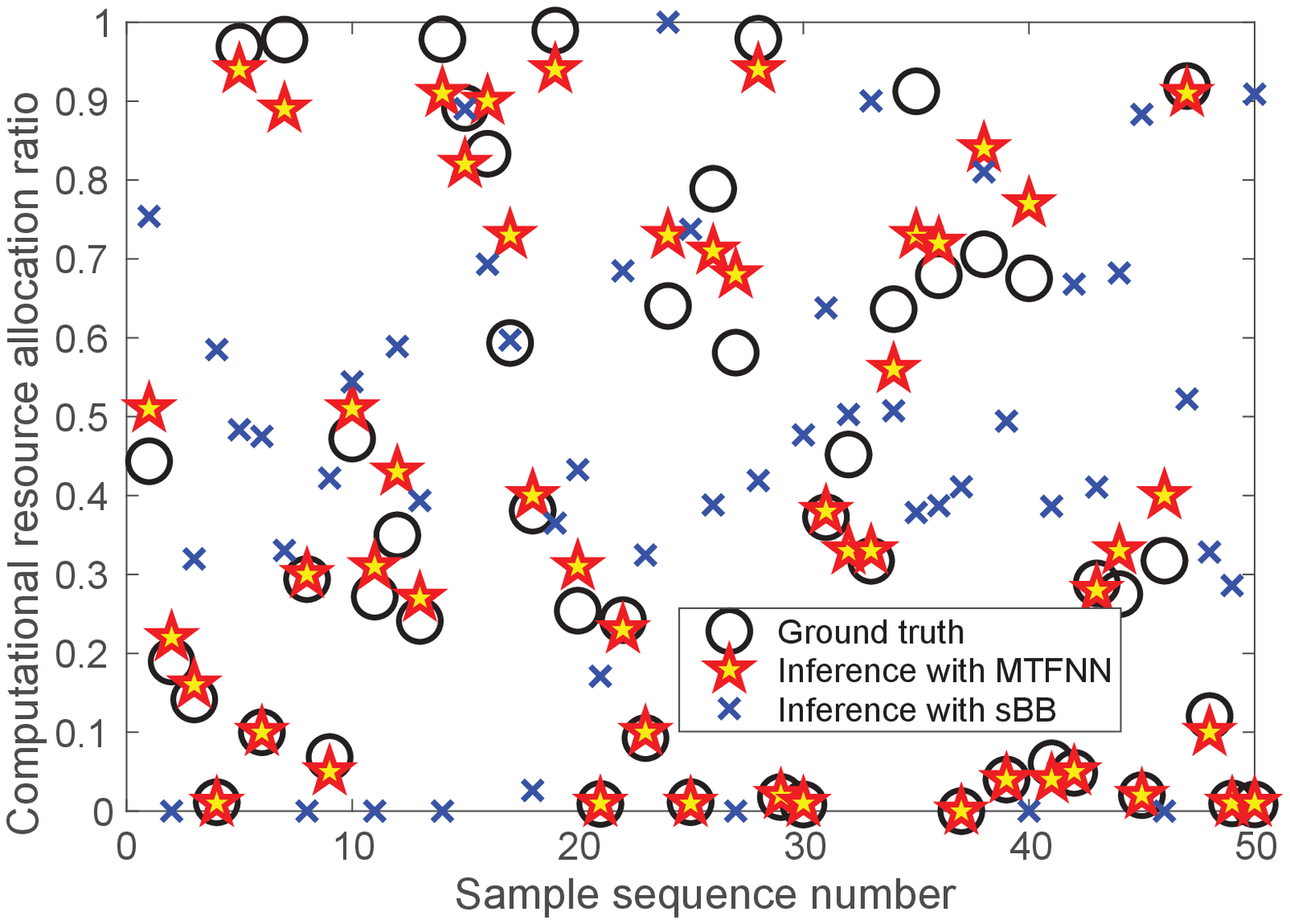}}
\caption{The predicted computational resource ratio of the MES (i.e., $\mathbf \Theta=[\Theta_1, \Theta_2, \Theta_3]$), where the number of devices is $3$. $\Theta_1$, $\Theta_2$ and $\Theta_3$ are respectively shown in (a), (b) and (c)   }
\label{regression}
\end{figure*} 

\subsection{Offline Training}
During the training phase, we train the MTFFN model which contains $2$ hidden layers, as shown in Fig \ref{fig02}, using the collected data.  In our MTFFN model, for the classification problem, the probability of each class is predicted using the Softmax function, i.e., the predicted probability for the $j$-th class given a sample vector ${\mathbf x}$ and a weighting vector ${\mathbf w}$ is $P(y=j|\mathbf{x})=\frac{{\rm exp}({{\mathbf x}^{\rm T}{\mathbf w}_j})}{\sum_{k=1}^K{\rm exp}({{\mathbf x}^{\rm T}{\mathbf w}_k})}$, where $K$ is the number of classes. We conventionally set the loss function of the multi-class classification (denoted as $\boldsymbol{l}_c$) as cross-entropy \cite{bce}. For the regression problem, the loss function (denoted as ${\boldsymbol{l}}_r$) is calculated using mean square error (MSE) \cite{MSE}. In our proposed MTFFN model, the loss function is defined as $\boldsymbol{l} = \chi_1 {\boldsymbol{l}_c}+\chi_2 {\boldsymbol{l}_r}$, where $\chi_1$ and $\chi_2$ are the weights. Here, we have $\chi_1=\chi_2=1$ and the Adam optimizer \cite{Adam} is used to optimize the MTFFN model. Therefore, the prediction of ${\mathbf{D}^*}$ and ${\mathbf{\Theta}^*}$ can be obtained when $\boldsymbol{l}$ is minimized. It should be noted that several methods have been proposed to scale up deep neural network (DNN) training across graphics processing unit (GPU) clusters\cite{GPU}, which helps to reduce the runtime of the offline training.

\section{Results and Discussions}
\subsection{Testing Results}
During the testing phase, the performance of the MTFFN model is evaluated based on the outputs\footnote{The outputs obtained from the MTFFN model are performed $50$ epochs and normalized to make sure the condition $\mathbf{C4}$ is met.} and the corresponding labels. To demonstrate the superiority on resolving the MINLP problem using the proposed MTFFN model, we compare with a benchmark scheme ``sBB" which is implemented using the MATLAB toolbox of the APMonitor Optimization Suite\cite{APmonitor}. 
 
\subsubsection{Inference accuracy}
The inference accuracy of getting $\mathbf{D}^*$ and $\mathbf{\Theta}^*$ are defined as follows. We define $\eta=\frac{Number \ of \ correct\  predictions}{Total \ number \ of \ predictions}$ to indicate the accuracy of the offloading decision making (a.k.a. multiclass classification). We use the MSE to indicate the accuracy of resource allocation strategy (a.k.a. regression), i.e., $\varepsilon=\frac{1}{mN}\sum_{i=1}^m\sum_{j=1}^N(y_j^i-x_j^i)^2$, where $m$ denotes the total number of samples, $N$ is the total number of devices. $y_j^i$ is the predicted value of $d_j$ from the $i$-th sample and $x_j^i$ is its label.

\begin{table}[t]  
\renewcommand{\arraystretch}{1.2}
  \captionsetup{font={small }} 
\caption{\\ \scshape Computation Accuracy and Complexity } 
\label{notations}  
\centering  
\begin{tabular}{c|c|c}
 \hline
 \diagbox{$N$}{$\eta$, $\varepsilon$, $t$}{$S$} 
 & sBB & MTFFN\\
 \hline
 2
 & $70\%, \ 0.055, \ 14.1 \ ms$ & $96\%, \ 0.016, \ 2.5 \ \mu s$ \\
 3 
 & $62\%, \ 0.047, \ 14.2 \ ms$  & $89\%, \ 0.027, \ 2.5 \ \mu s$ \\
  4  
  & $58\%, \ 0.053, \ 14.5 \ ms$ & $83\%, \ 0.029, \ 2.2 \ \mu s $  \\
 \hline
 \end{tabular}  
\end{table} 

\subsubsection{Computation complexity}
In this paper, the computation complexity denotes the execution time per sample, which is defined as $t=\frac{Total \ execution \ time}{Number \ of \ samples}$. As the number of devices (denoted as $N$) grows, the conventional exhaustive search strategy suffers from the exponential time complexity $O((2g)^N)$, where $g=\frac{1}{\omega}+1$, $\omega \in (0, 1)$ denotes the granularity of computational resource allocation. Meanwhile, to solve the MINLP problem, the sBB always has exponential worst-case complexity, i.e., $O(2^N)$\cite{BB}. In our proposed MTFFN model, the quadratic time complexity can be achieved as $O(M^2L)$, where $L$ is the number of layers, $M$ is the number of neurons in a hidden layer which indicates the scale of the neuron network model. Moreover, we just need to train our learning model once, which can be performed offline via the machines with strong computing and storage capabilities, e.g., the GPU clusters. Therefore, the MTFFN model has a relatively low complexity compared to the ``sBB" and exhaustive search schemes.

The inference accuracy ($\eta$ and $\varepsilon$) and computation complexity ($t$) are reported in Table II\footnote{Note that the ``Schemes" is abbreviated as ``$S$" to save space in Table II.}. It can be observed that compared to the ``sBB", our proposed ``MTFNN" model obtains the much lower complexity on the premise of a relatively high accuracy, i.e., the time cost of ``MTFNN" is less than one-tenth of one percent of the ``sBB", and outperforms the ``sBB" by average $40\%$ in the classification inference accuracy. Moreover, the performance comparison of regression inference for the case of three devices is presented in Fig. \ref{regression}. It can be observed that the proposed ``MTFNN" model predicts the computational resource ratio more accurately, which match the ground truth well.

\begin{figure}[t]
  \centering
  \captionsetup{font={small }}
  \includegraphics[width=4.0in, height=2.5in]{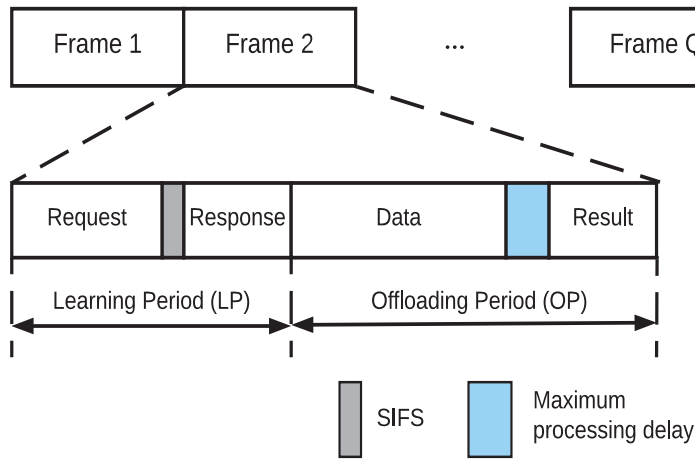}
  \caption{MAC framework}
  \label{fig03}
\end{figure}
%
%
\subsection{Implementation}
A MAC framework is presented to achieve channel coordination and intelligent offloading, as shown in Fig. \ref{fig03}, where the MTFNN module can be deployed onto the MES co-located with the AP \cite{VTC}. Different from using TDMA and OFDMA protocols for the uplink offloading in \cite{TDMA} and \cite{OFDMA}, in our MAC framework, the time is divided into $Q$ frames, each of which is subdivided into a learning period (LP) and an offloading period (OP). During the LP, all devices with tasks to be processed send a ``Request" message to AP using NOMA\footnote{Although an uplink NOMA scenario is assumed in this paper, our proposed MTFFN model can be also extended into other orthogonal multiple access schemes with a minor modification on the communication model, and then re-train the model offline.}, which contains the input parameters to the MTFFN model. On receiving the ``Request", AP decodes the ``Request" message based on the SIC technology and then predicts  $\mathbf{D}^*$ and $\mathbf{\Theta}^*$ based on the MTFNN model\footnote{We assume that the channel remains static within each frame.}. Then, AP notifies the devices of the predicted $\mathbf{D}^*$ and $\mathbf{\Theta}^*$ by replying a ``Response" message. During the OP, the devices with $\mathbf{D}^* \neq 0$ send the data needed to be processed to AP simultaneously in a NOMA way. On receiving the data, AP processes the corresponding tasks with the computational resources allocated to these tasks based on $\mathbf{\Theta}^*$, i.e, $\mathbf{F}^* = \mathbf{\Theta}^* \cdot F$. After all the tasks have been processed, AP returns the executed results back to the devices.

\section{Related Works}
Recently, joint optimization of computation offloading strategy and computing resources allocation to achieve different objectives has received ever-increasing attention\cite{{Communicating while computing},{Guo},{ROF},{Eom},{Yu}}. Specifically, a large body of existing works solve the optimization by seeking an optimal or sub-optimal solution using mathematical algorithms. For example, the authors in \cite{Communicating while computing} proposed an optimal offloading strategy using convex optimization. In \cite{Guo}, an efficient game-theoretic computation offloading scheme was proposed for MEC in 5G HetNets. However, due to the time-varing characteristic of the wireless channel, the previous works need to resolve the optimization problem very frequently to obtain the optimal/sub-optimal offloading decision, which introduces a large overhead to the MEC system. To tackle this problem, combining machine learning with the computation offloading has become an effective and attractive solution. In \cite{ROF}, an optimal offloading scheme was proposed for intermittently connected fog system using Markov decision process (MDP). In \cite{Eom}, a machine learning-based runtime adaptive scheduler was proposed for a mobile offloading framework based on past behavior and current conditions. In \cite{Yu}, an offloading decision problem was formulated as a multi-label classification problem for a single-user single-cell scenario, and a deep supervised learning method is developed to minimize the system overhead. Even though the offloading decision-making problem can be solved in \cite{{ROF},{Eom},{Yu}}, the computational resource allocation problem for the resource-limited MES is not considered. 

\section{Conclusions}
In this paper, we presented a multi-task learning (MTL) enabled offloading framework for MEC networks with uplink NOMA. We first formulated the joint optimization problem of offloading decision and MES computation resource allocation as a mixed integer nonlinear programming problem. Then, we developed an MTL based feedforward neural network model to solve the optimization problem more efficiently with high accuracy. Simulation results demonstrate that our proposed offloading approach achieves better performance than other benchmarks in terms of system cost saving. This paper is one of our first attempts to integrate MEC system design with machine learning. Future work is in progress to take more complicated scenario into consideration.

\vspace{12pt}

\end{spacing}
\end{document}